# Motion-Informed Deep Learning for Brain MR Image Reconstruction Framework

Zhifeng Chen[1, 2, #], Kamlesh Pawar[1 #], Kh Tohidul Islam[1], Himashi Peiris[2], Gary Egan[1], and Zhaolin Chen[1, 2 *]

[1] Monash Biomedical Imaging, Monash University, Clayton, VIC, Australia
[2] Department of Data Science & AI, Faculty of IT, Monash University, Clayton, VIC, Australia

**Address correspondence to:**
Zhifeng Chen, zhifeng.chen@monash.edu;
Zhaolin Chen, zhaolin.chen@monash.edu;
Monash Biomedical Imaging, 770 Blackburn Road, Monash University, Clayton, VIC 3168, Australia

Zhifeng Chen and Kamlesh Pawar contributed equally to this work and shared the first co-authorship.

# Abstract

Motion artifacts in Magnetic Resonance Imaging (MRI) are one of the frequently occurring artifacts due to patient movements during scanning. Motion is estimated to be present in approximately 30% of clinical MRI scans; however, motion has not been explicitly modeled within deep learning image reconstruction models. Deep learning (DL) algorithms have been demonstrated to be effective for both the image reconstruction task and the motion correction task, but the two tasks are considered separately. The image reconstruction task involves removing undersampling artifacts such as noise and aliasing artifacts, whereas motion correction involves removing artifacts including blurring, ghosting, and ringing. In this work, we propose a novel method to simultaneously accelerate imaging and correct motion. This is achieved by integrating a motion module into the deep learning-based MRI reconstruction process, enabling real-time detection and correction of motion. We model motion as a tightly integrated auxiliary layer in the deep learning model during training, making the deep learning model 'motion-informed'. During inference, image reconstruction is performed from undersampled raw k-space data using a trained motion-informed DL model. Experimental results demonstrate that the proposed motion-informed deep learning image reconstruction network outperformed the conventional image reconstruction network for motion-degraded MRI datasets.

# 1. INTRODUCTION

Motion is one of the most challenging problems in MRI due to its sporadic occurrences. In brain MRI, the movements caused by voluntary or involuntary motions often result in motion artifacts which include blurring on edges or loss of details of anatomical structures. Patient motion carries significant costs to healthcare providers, it is estimated that almost 30% of MRI clinical examinations suffer from motion issues resulting in repeated scans and overhead costs for nurses, radiographers, and radiologists [1]. Motion can significantly degrade image quality that requires repeating gadolinium contrast administration, and inconvenience to the patient and their caregiver. The goal of reducing motion artifacts through improved imaging strategies represents an opportunity to improve the quality and efficiency of MR imaging services, ultimately improving the delivery of healthcare services.

Researchers have explored numerous solutions to solve the motion issue in MRI [2]–[9]. The most straightforward way to deal with motion is reacquisition, however it lengthens the scan time. Other approaches to tackling brain motion include the use of external hardware such as an optical tracking system [2] and active marker headband [3], navigator- and image-based motion tracking [4], prospective motion correction [2]–[4], and retrospective motion correction [4]. In addition, non-Cartesian imaging, including radial [5], propeller [6], and spiral [7], [10] acquisitions provide motion robustness solutions to obtain motion-free MR images. Moreover, there are also post-processing techniques like data rejection and clustering-based approaches [8], [9]. Xu et al. [9] synergistically include motion-informed registration-based motion correction in reconstruction to improve motion robustness. Motion sorting strategy has also become popular in recent years [11]–[16]. Additionally, researchers propose to use structured low rank to remove physiological motion in the brain [17], [18].

Recent studies have shown promising results of using deep learning for MRI motion correction [19]–[26]. Pawar et al. [24] proposed the InceptionResNet, which employed a linear combination of rigidly transformed data and clear images. Training the InceptionResNet network with these motion-simulated images resulted in performance surpassing that of the entropy minimization method. Deep learning is also widely used in undersampled MRI reconstruction [27]–[33]. However, there are very few integrated reconstruction methods that can detect and correct motion for undersampled MRI data.

In this study, a novel motion detection and correction incorporated image reconstruction framework is proposed. Within this framework, a motion detection module is seamlessly integrated into the motion-informed variational network image reconstruction model for undersampled MRI data. We have performed experiments to validate the proposed framework using both undersampled retrospectively simulated motion data and prospectively acquired motion datasets.

In the following section 2, we begin by elucidating the fundamental problem statement, followed by a thorough examination of each module related to the reconstruction framework. We then present a comprehensive model for the motion-informed MRI reconstruction scheme. Section 3 serves to exemplify the efficacy of our proposed scheme through the presentation of findings from simulated fastMRI experiments and several *in vivo* brain imaging experiments. Section 4 is dedicated to a comprehensive discussion of the advantages and limitations inherent in the proposed methodology. Finally, we outline prospective avenues for further research.

## 2. Methods

This section overviews each component within the motion detection, image reconstruction, and motion correction processes. Following this, we delve into an in-depth exploration of each module, encompassing the support vector machine (SVM) for motion detection, the motion simulation layer, the motion-informed data consistency parameter (MIDCP) estimator, and the deep learning image reconstruction model.

### 2.1. Framework overview

The illustrated framework for motion detection, image reconstruction, and motion correction is presented in Figure 1. To begin, a motion simulation module is applied to simulate motion corrupted k-space, followed by a motion detection module using SVM on the initial reconstruction. During the motion detection phase, the k-space data affected by motion is fed into the motion-informed reconstruction model VarnetMi (motion-informed variational network) for simultaneous motion correction and image reconstruction. Subsequently, the motion-free image reconstruction results are generated. Additionally, the motion detection module outputs the motion status ('Yes' or 'No' in Figure 1).

#### 2.1.1. Motion simulation layer

The motion simulation layer simulates motion during model training phase and generates random motion parameters in three degrees of freedom, two translation parameters with a maximum of +/- 10 pixels, and one rotation parameter with a maximum of +/- 10 degrees. The number of motion events for each image varied from 0 to 16 i.e., a set of three motion parameters were generated randomly up to a maximum of 16 times, and the k-space was distorted using these motion parameters. The motion layer was incorporated in the motion-informed variational network as shown in Figure 2.

**2.1.2. Motion-informed image reconstruction model: VarnetMi**

Our reconstruction model is built upon an end-to-end variational network [34] consisting of a cascade of convolutional neural network (CNN) driven reconstruction steps described as follows:

$$k^{t+1} = k^t - H^t(k^t)\Omega(k^t - \hat{k}) + R^t(k^t) \tag{1}$$

where, $k^t$ is the current k-space data, $k^{t+1}$ is the updated k-space data, $H^t$ is a CNN (Figure 2b) that takes k-space and predicts a single parameter for data consistency. $\hat{k}$ is the acquired data, $\Omega$ is the mask of sampling locations, and $R^t$ is the reconstruction CNN (Figure 2e). Eq.1 is equivalent to one step of gradient descent. The reconstruction CNN $R^t$ proceeds as follows: (i) uses intermediate k-space, (ii) performs the inverse Fourier transform, (iii) combines the multi-channel images to a complex-valued image using the sensitivity maps estimated from the sensitivity maps estimation (SME) network (Figure 2f), (iv) processes the combined complex-valued image through a Unet, (v) converts the processed image back to multi-channel k-space and (vi) enforces data consistency (DC) (Figure 2c). To make the variational network 'motion-informed' we incorporate a motion simulation component (Figure 2d) and motion-informed data consistency parameter estimator (MIDCP) (Figure 2b). After the last iteration, the k-space was converted to the root of the sum of squares (rss) of the image (Figure 2a) which was used to calculate the loss for supervised training and the loss function was the difference in structural similarity [35].

It is worthwhile to note that the output of $H^t$ will simply converge to a constant value for the non-motion-informed variational network [34]. However, in the proposed motion-informed DL model, the output of $H^t(k^t)$ depends on the input k-space data and thus the output of $H^t(k^t)$ will be different for each input data. This output signal can be further used to detect motion using classification models. To achieve this, we utilized a support vector machine to detect motion occurrence in the data.

2.2. Support vector machine for motion detection

SVM is an efficient and accurate supervised classification technique [36], that discerns data patterns between two groups through a supervised training session. In the model, SVM is applied to classify input data to either motion corrupted or motion free during training and validation. The SVM is trained based on the initial reconstruction following MIDCP as shown in Figure 2b. Each MIDCP produces one value, which serves as input to the SVM module for determining the presence or absence of motion. The output indicates the motion status of the input data, with "yes" indicating motion detected and "no" indicating no motion. The training method [36] was applied separately and integrated with the image reconstruction network to determine the motion status of the k-space data.

## 3. EXPERIMENTS

### 3.1. Experiments using fastMRI data

The brain images from the fastMRI dataset were used for a benchmark that contains T1-weighted (some with post-contrast), T2-weighted, and fluid-attenuated inversion recovery (FLAIR) images. For detailed information of all datasets, please reference [37] regarding the method of undersampling, with an equispaced mask adopted as the sampling pattern. Regarding the data partitioning, we first separated all multi-slice volumes into training and testing groups and then split the volume into slices (i.e., images). Training images were reconstructed from multi-channel k-space data with undersampling, and after coil combination the complex image datasets were scaled to a magnitude range of $[-1, 1]$ without loss of phase information. The coil sensitivity maps were calculated with the sensitivity maps estimation (SME) network [34]. During the training, real and imaginary parts of all images were separated into two real and imaginary channels when input into the neural network. 1300 images of size $320 \times 320$ from the dataset used in [37] were used to train Varnet and VarnetMi. 1000 images were used for training, and 300 images were used for testing.

### 3.2. Experiments in prospective motion corrupted data

Motion correction experiments were performed under Monash University institutional ethics approval for healthy participants. Two separate scans were performed for each volunteer: one without motion, where the volunteers were instructed to remain still, and a second with motion, where the volunteers were allowed to move naturally during the scan. All the subjects signed the informed consent before the imaging experiments at Monash Biomedical Imaging.

Axial 2D T1-weighted turbo spin-echo (TSE) brain datasets were acquired on a 2D axial TSE on a 3 T MRI scanner (Skyra, Siemens Healthineers GmbH, Erlangen, Germany) and using a T1-weighted imaging contrast. The gradient performance of the scanner was a gradient strength of 45 mT/m and a slew rate of 200 T/m/s. The imaging parameters included: repetition time (TR)/inversion time (TI)/echo time (TE) = 580/100/7.9 ms, matrix size = 320 × 384, and number of coils = 32. Readout oversampling factor 2, field of view 220 × 220 mm$^2$, slice thickness 1 mm, number of slices 24, and flip angle 150 degree.

Axial 2D T2-weighted TSE brain datasets were also acquired on the same scanner. The relevant imaging parameters included: TR/ TI/ TE= 6000/100/103 ms, matrix size = 320 × 255, and number of channels = 32. Readout oversampling factor 2, field of view 220 × 220 mm$^2$, slice thickness 1 mm, number of slices 24, and flip angle 150 degree.

### 3.3. Image Reconstruction

The proposed framework was implemented in Python (Python Software Foundation, https://www.python.org/) with the deep learning framework PyTorch on an NVIDIA A40 GPU. All the data was reconstructed offline using a Linux (Ubuntu 20.04.6 LTS) server with 52-Core Intel(R) Xeon(R) Gold 5320 CPU @ 2.20GHz and 1 TB of memory.

### 3.4. Evaluation Criteria

The normalized mean square error (NMSE) was used to evaluate the reconstructed results. The NMSE is defined as the following:

$$NMSE = \frac{\|I_{ref} - I\|_F}{\|I_{ref}\|_F} \qquad (2)$$

where $I_{ref}$ stands for the reference motion-free image (fully sampled reconstruction), and $r$ represents the spatial locations of the image. $I$ is the under-sampled motion corrupted reconstruction result. $F$ represents the Frobenius norm.

To provide a further quantitative evaluation of the reconstructed images, the peak signal-to-noise ratio (PSNR) and structural similarity index (SSIM) were adopted [38]. PSNR is commonly utilized to measure the reconstruction quality by comparing the benchmark ($I_{ref}$) with the reconstructed result, leveraging the fully sampled references available in the cardiac simulation. On the other hand, SSIM assesses the similarity between the benchmark and the reconstructions, taking into consideration the characteristics of human visual perception and thus providing a more reliable measure. These two metrics are calculated as the following equations:

$$PSNR = 20(MAX_{I_{ref}(r)}/MSE) \tag{3}$$

$$MSE = \sqrt{\frac{1}{N}\sum_{r=1}^{N}[I_{ref}(r) - I(r)]^2} \tag{4}$$

$$SSIM(I_{ref}, I) = \frac{(2\mu_{I_{ref}}\mu_I + c_1)(2\sigma_{I_{ref}I} + c_2)}{\left(\mu_{I_{ref}}^2 + \mu_I^2 + c_1\right)\left(\sigma_{I_{ref}}^2 + \sigma_I^2 + c_2\right)} \tag{5}$$

Here $I_{ref}$ and $I$ and $r$ are the same as the previous. $MAX_{I_{ref}}$ is the maximum signal intensity of $I_{ref}$. MSE represents the mean square error of the reconstructed image. The mean signal values of $I_{ref}$ and $I$ are $\mu_{I_{ref}}$ and $\mu_I$. The variance of $I_{ref}$ and $I$ are $\sigma_{I_{ref}}^2$ and $\sigma_I^2$. The covariance of $I_{ref}$ and $I$ is $\sigma_{I_{ref}I}$, $c_1$ and $c_2$ are two variables to stabilize the equation when the denominator is too small.

To evaluate motion detection performance from the SVM module, we calculate true positives (TP), true negatives (TN), false positives (FP), and false negatives (FN), as detailed in Table 1. Additionally, we calculate the Accuracy, Precision, Sensitivity and Specificity based on the previous four given values, as the following:

$$Accuracy = (TP + TN)/(TP + TN + FP + FN) \tag{6}$$

$$Precision = TP/(TP + FP) \tag{7}$$

$$Sensitivity = TP/(TP + FN) \tag{8}$$

$$Specificity = TN/(TN + FP) \tag{9}$$

### 3.5. Ablation Study

To gain deeper insight into the effectiveness of our proposed motion-informed deep learning image reconstruction method, we conducted an ablation study with fastMRI data. The aim was to assess the performance contribution of VarnetMi when applied to motion-free data. This analysis not only allowed us to evaluate the impact of motion correction but also provided valuable insights into the overall efficacy of our reconstruction approach under different conditions. By systematically evaluating the performance of VarnetMi in scenarios with and without motion, we aimed to better understand its role in enhancing image quality across a range of imaging conditions.

## 4. RESULTS
### 4.1. Results from fastMRI data

The study used fastMRI data involved experiments with both simulated motion and motionless images, and included various MR contrasts (T1, T2, T1 post-contrast, and FLAIR). We compared the performance of Varnet and VarnetMi with motion using a four-fold undersampling accelerated imaging. The results revealed a significant divergence in quantitative performance between the two algorithms with VarnetMi providing superior results in the presence of motion. Specifically, reconstructions produced by Varnet showcased noticeable motion artifacts, particularly visible in the zoomed-in subfigures presented in Figures 3 and 4. In contrast, VarnetMi, equipped with motion-informed capabilities, effectively restored images without ghosting or aliasing artifacts. The quantitative results were consistent with visual observation. In all scenarios of motion degraded data, the VarnetMi outperforms Varnet with lower NMSE, higher PSNR and SSIM, as indicated in Table 2.

The confusion matrix obtained after the network is trained and tested with test data is given in Table 1. The SVM module reported a 94.5% accuracy in detecting motion, including correctly classified 39419 test data (with true positive 19419, and true negative 20000) and mis-classified 2293 data (with false positive 1265, and false negative 1028). From the analysis of experimental results, it was observed that the proposed SVM motion detection system achieved a 94.5% accuracy rate (specificity/sensitivity/precision 95%/94%/95%).

### 4.2. Results from prospective motion data

Figure 5 presents the reconstructions of prospective motion corrupted T1-weighted and T2-weighted brain data. In this figure, we compare the results of the non-motion-informed Varnet reconstructions with those of the proposed motion-informed reconstruction scheme, VarnetMi, as compared to the fastMRI reconstruction results. A visual examination reveals that the proposed motion-informed VarnetMi scheme consistently outperforms the non-motion-informed Varnet approach, as evident in the image quality.

This observation was further validated by evaluating various metrics, including NMSE, PSNR, and SSIM in Table 3. The VarnetMi reconstructions consistently exhibited superior performance compared to Varnet, with lower NMSE, higher PSNR, and higher SSIM scores, underscoring the effectiveness of the motion-informed approach.

### 4.3. Ablation study results

In our comprehensive ablation study, we systematically compared the performance of Varnet and VarnetMi across various imaging modalities, including T1, T1 post-contrast, T2, and FLAIR images, specifically examining their outcomes in the absence of motion (as illustrated in Figure 6). The meticulous analysis revealed that Varnet exhibited slightly superior performance compared to VarnetMi under these conditions. Remarkably, in both sets of results—with and without motion-informed image reconstruction—no discernible artifacts were observed in the reconstructed images, as visually depicted in Figure 6. This robust performance across diverse imaging modalities and motion scenarios underscores the reliability and effectiveness of the VarnetMi approach in achieving artifact-free reconstructions.

Table 4 presents a comparison of reconstruction metrics between Varnet and VarnetMi for motion-free data, including NMSE, PSNR, and SSIM. Upon examination, several observations emerge. In instances where no motion was detected in the input images, the performance disparity between Varnet and VarnetMi was minimal, with VarnetMi exhibiting slightly inferior results. While slightly worse in NMSE, PSNR, and SSIM were noted across all examined imaging contrasts, the differences were negligible, as visually depicted in Figure 6.

Moreover, a comprehensive quantitative analysis involving 1300 image slices from various subjects was conducted for another ablation study, which compared Varnet with VarnetMi in two distinct scenarios encompassing various contrasts (as illustrated in Figure 7). Specifically, a version without motion detection was included in this examination. The results presented in Figure 7 demonstrate that the proposed motion-informed image reconstruction with SVM motion detection offers the best overall performance, characterized by superior NMSE, PSNR, and SSIM metrics. Notably, the exclusion of the motion detection module resulted in a slight decrease in the performance of the proposed framework, particularly when the correct image reconstruction module was not utilized in the subsequent reconstruction process.

In scenarios where motion was present, VarnetMi outperformed Varnet, exhibiting lower NMSE, higher PSNR, and correspondingly higher SSIM. Conversely, in cases where motion was absent, the performance of Varnet was marginally better than that of VarnetMi. However, this difference was negligible.

## 5. DISCUSSION

In this work, a motion correction-based image reconstruction framework was proposed for undersampled MRI. The effectiveness of the new method was illustrated by the fastMRI and *in vivo* data. The proposed VarnetMi network demonstrated superior performance in achieving high-quality reconstructions with motion degraded MRI. We introduced an innovative approach to infuse motion information into deep learning training, specifically designed to undersampled motion degraded MRI scans. Our methodology entailed the integration of motion modeling as a tightly integrated auxiliary layer within the deep learning model during the training phase, effectively endowing the deep learning model with a heightened sense of motion information. Consequently, during the inference stage, this motion-informed layer seamlessly came into play, facilitating direct motion corrected image reconstruction from the raw k-space data.

In this study, we also introduced an SVM motion detection module as a crucial pre-processing step, significantly enhancing the reliability of the proposed technology. Especially in the case of motion present data, the analysis consistently demonstrated that VarnetMi exhibited substantial improvements in terms of SSIM, PSNR, and NMSE, as depicted in Figures 3, 4 and 5. In contrast, it became evident that VarnetMi's performance is slightly compromised when applied to motion-free data. Thus, the integration of an SVM-based motion detection system is imperative, affording the proposed technique more versatility.

In prior experimental analyses, a distinct performance advantage of the motion-informed VarnetMi was evident in T2-weighted images as opposed to T1-weighted results. This observation underscores a substantial enhancement achieved by the proposed VarnetMi in the context of T2-weighted imaging when compared with the conventional Varnet. The observed disparity in performance can be attributed to the T2-weighted imaging modality's heightened sensitivity to motion when contrasted with T1-weighted imaging. This nuanced sensitivity to motion in T2-weighted images positions VarnetMi as a valuable solution for motion mitigation in the application of this specific imaging modality.

One limitation of the proposed motion-informed image reconstruction method is its restriction to rigid motion correction, including displacement and rotation movements. Currently, the technique relies on a framework designed for correcting rigid motion and the model has not been optimized for non-rigid motion. Rigid motion is the dominant motion component in brain imaging. In contrast, when it comes to abdominal imaging, including liver and cardiac imaging, non-rigid motion becomes an inevitable challenge that must be addressed during the training process. Another limitation is that the proposed motion-informed image reconstruction has only been tested on healthy subjects. To verify the robustness of the motion correction network, it is essential to include patients with pathology data in future studies. Additionally, it is worth noting that when motion is absent, the motion-informed image reconstruction yields slightly inferior results compared to the original non-motion-correction scheme. This aspect should be addressed in future research to enhance the applicability and effectiveness of the proposed technique.

It is important to highlight that there are still opportunities for further investigation. For instance, one potential avenue could explore the integration of other deep learning image reconstruction methods, like deep image prior [30], to enhance the effectiveness of image reconstruction. Additionally, future research could focus on optimizing the sampling schemes through non-Cartesian trajectories to improve the motion robustness during acquisition.

## 6. CONCLUSION

A robust deep learning reconstruction method, founded on a motion-informed variational network, tailored for accelerated imaging was developed. The proposed approach is versatile, as it can effectively operate on both motion-degraded and non-motion-degraded undersampled k-space data. This versatility renders it highly suitable for a wide range of clinical applications, ensuring rapid and robust MRI outcomes.

**Conflict of interest**

The authors declare no competing interest.

**Data availability statement**

In the spirit of reproducible research, codes and data that support the findings of this study are available from the corresponding author or the senior author upon reasonable request.

**Acknowledgments**

The authors acknowledge the support of the Australian Research Council (ARC) DP210101863, and the ARC Fellowship Program IM230100002, as well as an early career researcher seed grant from Faculty of Information Technology, Monash University.

**Figure Captions:**

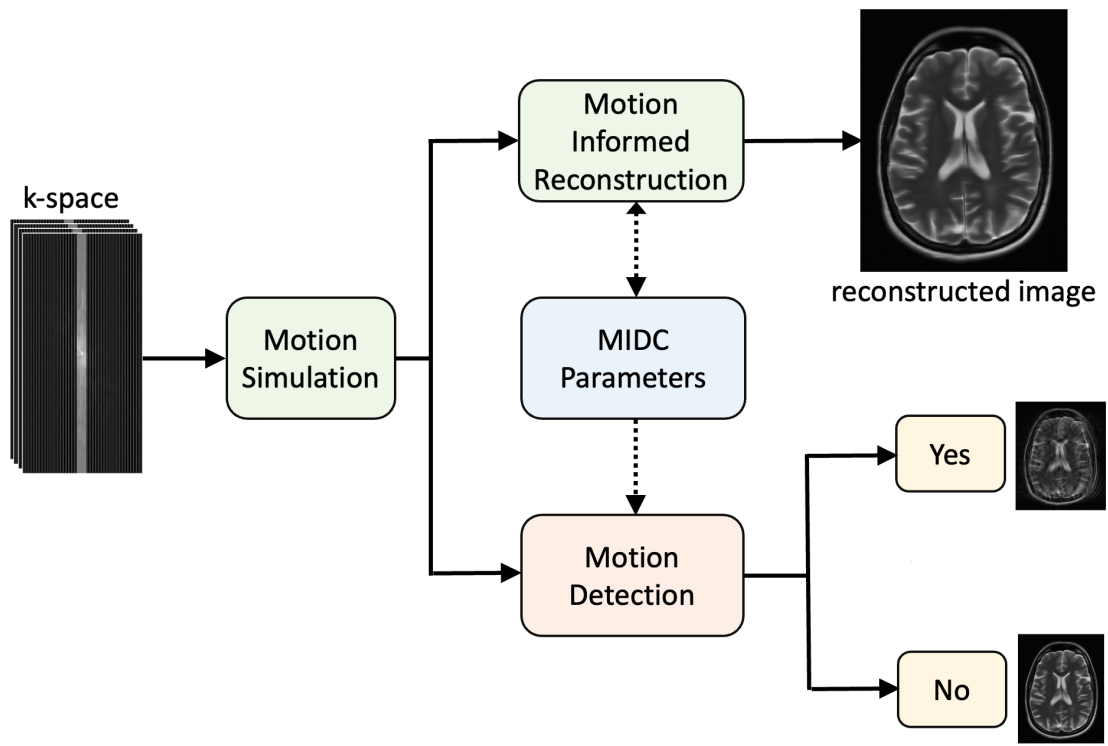

**Figure 1.** Overview of the proposed motion-informed reconstruction and detection framework. The motion-informed reconstruction includes simultaneous motion correction and Varnet image reconstruction; the motion detection module contains a support vector machine for motion detection. MIDC: motion-informed data consistency.

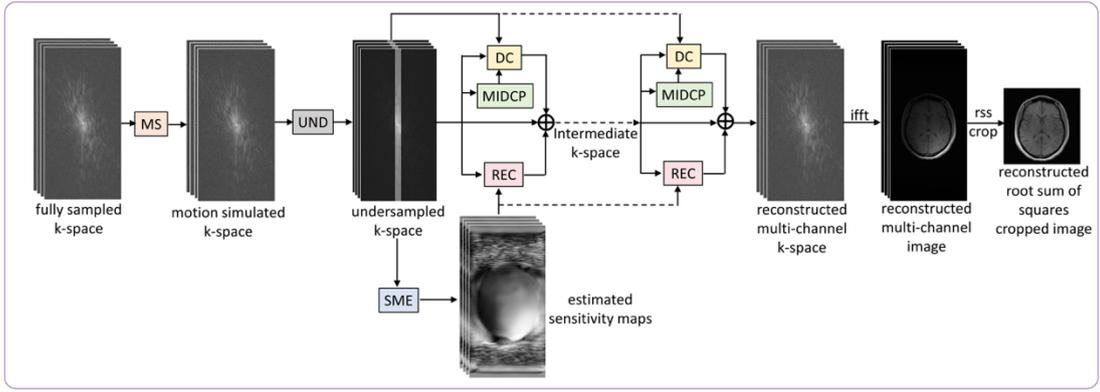

(a) Motion informed variational networks for accelerated imaging

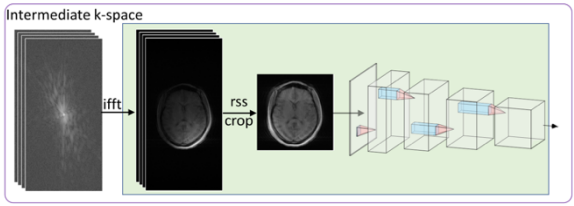

(b) MIDCP: motion informed data consistency parameter estimator

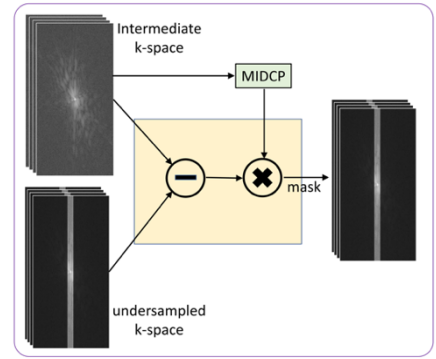

(c) DC: data consistency module

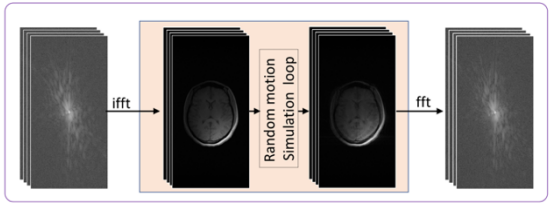

(d) MS: motion simulation module

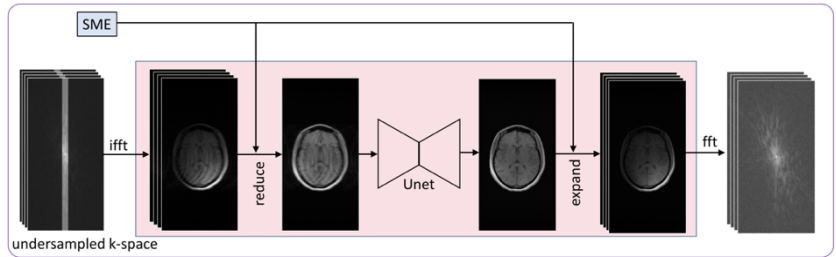

(e) REC: image reconstruction module

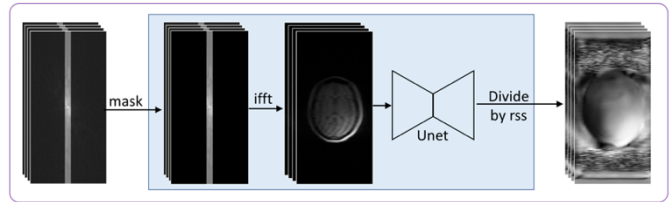

(f) SME: sensitivity map estimation module

**Figure 2.** (a) Block diagram of the proposed VarnetMi; (b) motion-informed data consistency module to estimate the appropriate data consistency parameter of data consistency (DC) module; (c) DC module enforcing the reconstruction to be consistent with the acquired data; (d) motion simulation module to simulate motion artifact for multi-channel k-space; (e) reconstruction module combines multi-channel k-space data to single complex-valued image and process it through a Unet; and (f) SME module that takes the center of k-space to estimate the sensitivity maps needed for reconstruction. UND: undersampled; rss: root of the sum of squares.

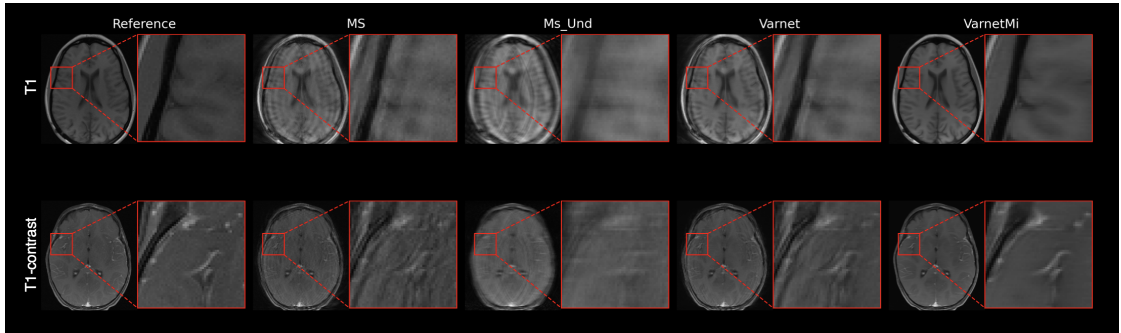

**Figure 3.** Representative images comparing Varnet and motion-informed VarnetMi in the presence of motion; Top row: T1-weighted images; Bottom row: T1 post-contrast weighted images; **Reference:** Ground truth images; **MS:** images corrupted with motion artifacts; **Ms_Und:** motion corrupted image undersampled with an acceleration factor of four; **Varnet:** reconstruction using the variational network; **VarnetMi:** reconstructions using the motion-informed variational network.

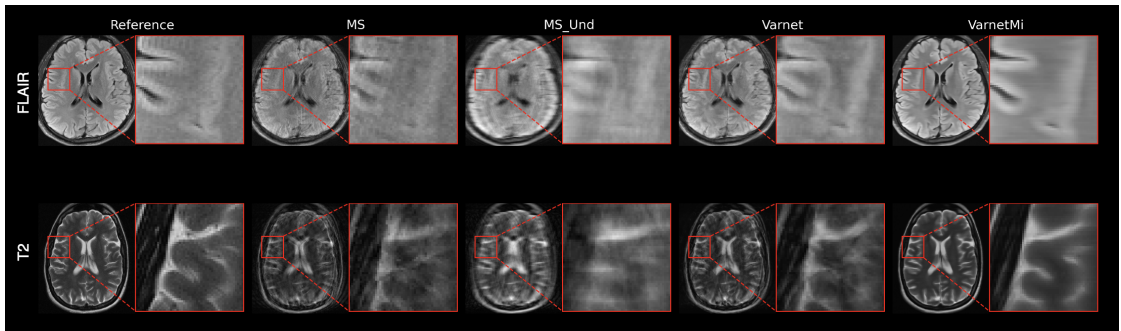

**Figure 4.** Representative images comparing Varnet and motion-informed VarnetMi in the presence of motion; Top row: FLAIR images; Bottom row: T2-weighted images; **Reference:** Ground truth images; **MS:** images corrupted with motion artifacts; **Ms_Und:** motion corrupted image undersampled with an acceleration factor of four; **Varnet:** reconstruction using the variational network; **VarnetMi:** reconstructions using the motion-informed variational network.

Table 1. Distinguishing attributes of normalized classification results based on predictions and ground truth.

| True label | Predict label | | |
|---|---|---|---|
| | | With motion | No motion |
| | With motion | 0.94 (True Positive) | 0.06 (False Negative) |
| | No motion | 0.05 (False Positive) | 0.95 (True Negative) |

Table 2. Results of VarnetMi vs. Varnet using motion-degraded fastMRI data.

| | | NMSE (%) | PSNR (dB) | SSIM (%) |
|---|---|---|---|---|
| T1 | Varnet | 5.37 | 27.91 | 84.58 |
| | VarnetMi | **0.72** | **36.61** | **95.66** |
| T1 post-contrast | Varnet | 2.36 | 30.93 | 86.18 |
| | VarnetMi | **1.39** | **33.25** | **90.85** |
| T2 | Varnet | 11.21 | 22.66 | 70.13 |
| | VarnetMi | **3.39** | **27.85** | **86.74** |
| FLAIR | Varnet | 2.21 | 29.75 | 82.60 |
| | VarnetMi | **0.98** | **33.30** | **90.51** |

Note: The lowest mean NMSE and highest mean PSNR/SSIM values are bold faced.

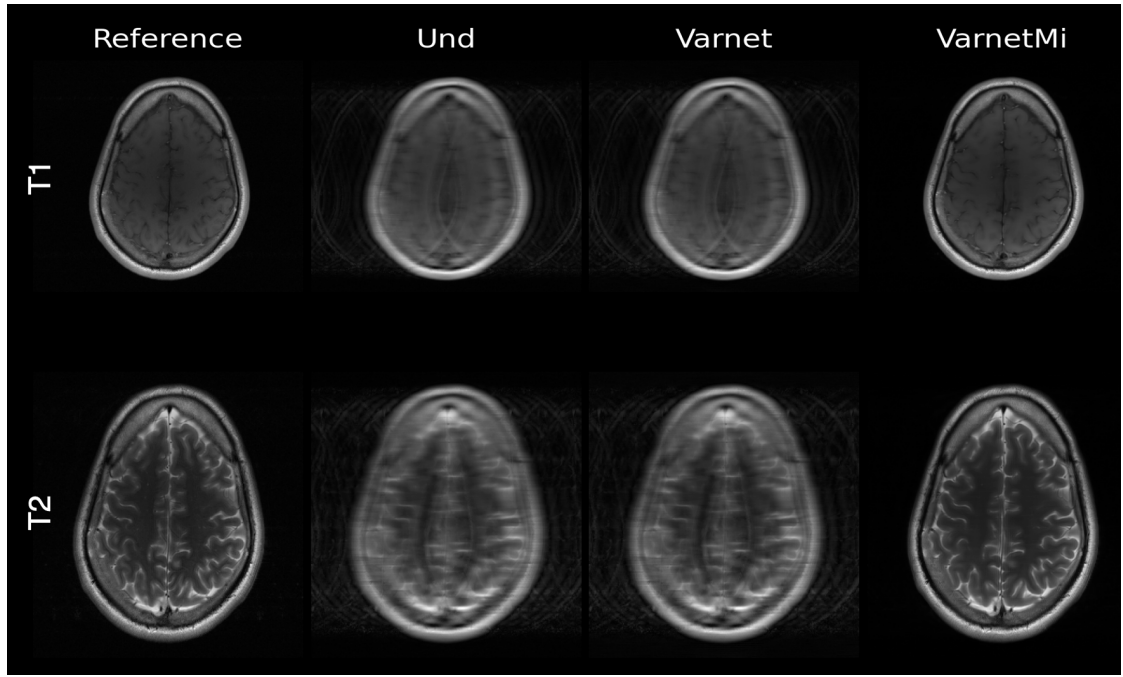

**Figure 5.** Representative images from healthy volunteer comparing Varnet and motion-informed VarnetMi in the presence of motion for *in vivo* data; Top row: T1-weighted images; Bottom row: T2-weighted images; **Reference:** Ground truth images; **Und:** motion corrupted image undersampled with an acceleration factor of four; **Varnet:** reconstruction using the variational network; **VarnetMi:** reconstructions using the motion-informed variational network.

Table 3. Results of VarnetMi vs. Varnet using motion corrupted in vivo data..

| | | NMSE (%) | PSNR (dB) | SSIM (%) |
|---|---|---|---|---|
| T1 | Varnet | 6.32 | 29.24 | 82.67 |
| | VarnetMi | **0.67** | **38.99** | **94.82** |
| T2 | Varnet | 7.13 | 26.99 | 78.28 |
| | VarnetMi | **0.46** | **38.84** | **97.95** |

Note: The lowest mean NMSE and highest mean PSNR/SSIM values are bold faced.

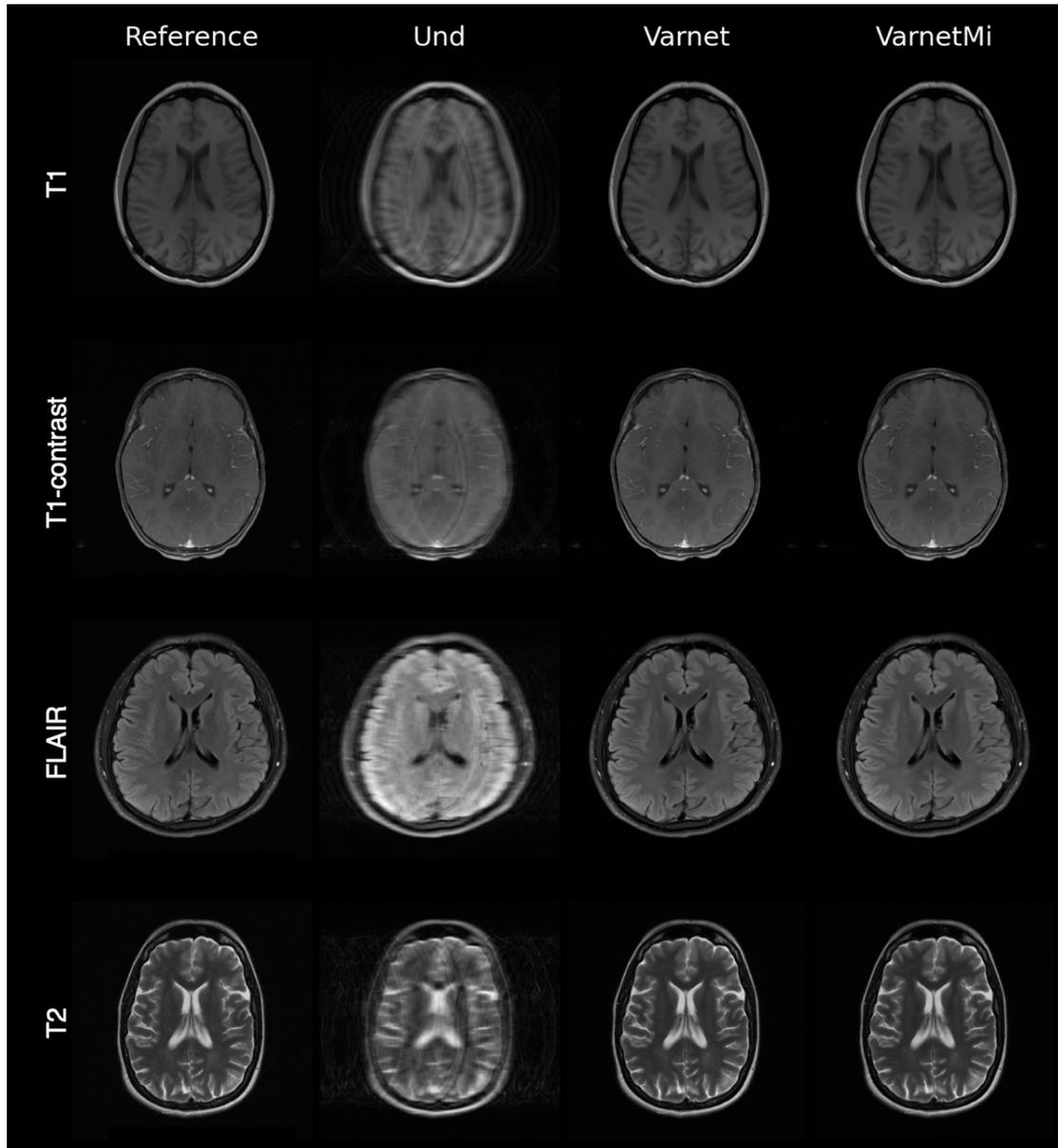

**Figure 6.** Representative images comparing Varnet and motion-informed VarnetMi in the absence of motion; **Reference:** Ground truth images; **Und:** image undersampled with an acceleration factor of four; **Varnet:** reconstruction using the variational network; **VarnetMi:** reconstructions using the motion-informed variational network.

Table 4. Different evaluation criteria for the reconstructed results of motion free data in Figure 5 with and without motion correction.

|  |  | NMSE (%) | PSNR (dB) | SSIM (%) |
|---|---|---|---|---|
| T1 | Varnet | **0.16** | **43.12** | **97.58** |
|  | VarnetMi | 0.19 | 42.29 | 97.39 |
| T1 post-contrast | Varnet | **0.30** | **39.86** | **94.71** |
|  | VarnetMi | 0.38 | 38.89 | 94.44 |
| T2 | Varnet | **0.48** | **36.35** | **93.50** |
|  | VarnetMi | 0.56 | 35.64 | 93.19 |
| FLAIR | Varnet | **0.23** | **39.57** | **95.85** |
|  | VarnetMi | 0.28 | 38.80 | 95.46 |

Note: The lowest mean NMSE and highest mean PSNR/SSIM values are bold faced.

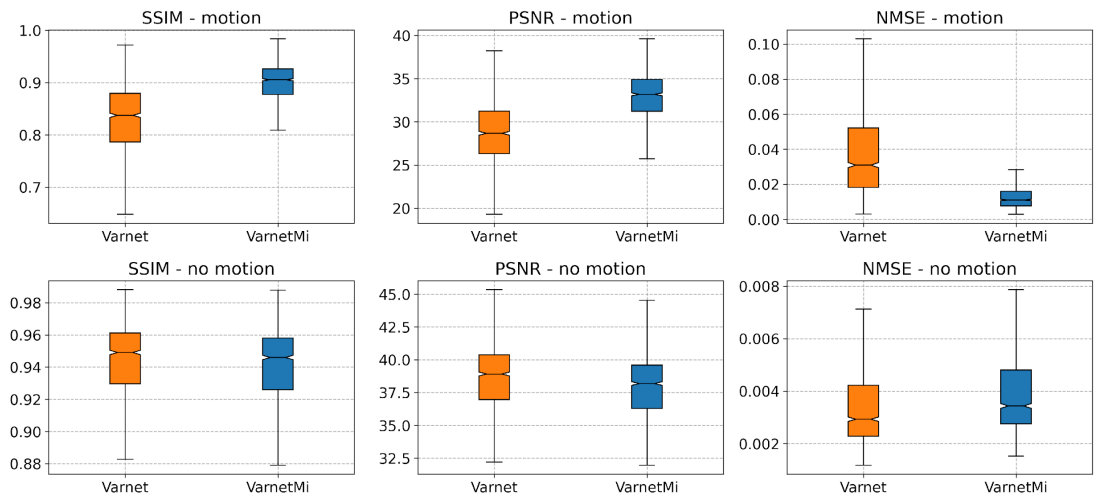

**Figure 7.** Group results on the 1300 slices (different subjects) consisting of T1, T2, T1 post-contrast, and FLAIR images. **Top row:** Quantitative scores for the reconstruction in the presence of motion; **Bottom row:** Quantitative scores for the reconstruction in the absence of motion. There was a marked improvement in the SSIM, PSNR, and NMSE using the proposed VarnetMi method compared to the Varnet method. In the absence of motion, there was a slight degradation in the performance of VarnetMi compared to Varnet.